\documentclass[preprint,showpacs,preprintnumbers,amsmath,amssymb]{revtex4}
\usepackage{epsfig}
\usepackage{graphicx}
\usepackage{dcolumn}
\usepackage{bm}

\newcommand{\ord}{{\cal O}}
\def\beq{\begin{equation}}
\def\eeq{\end{equation}}
\def\eeqn{\end{equation}}
\newcommand\iden{\leavevmode\hbox{\small1\normalsize\kern-.33em1}}

\newcommand{\bea} {\begin{eqnarray}}
\newcommand{\eea} {\end{eqnarray}}

\newcommand{\Gm}{\Gamma}

\let\jnfont=\rm
\def\NPB#1,{{\jnfont Nucl.\ Phys.\ B }{\bf #1},}
\def\PLB#1,{{\jnfont Phys.\ Lett.\ B }{\bf #1},}
\def\EPJC#1,{{\jnfont Eur.\ Phys.\ Jour.\ C }{\bf #1},}
\def\PRD#1,{{\jnfont Phys.\ Rev.\ D }{\bf #1},}
\def\PRL#1,{{\jnfont Phys.\ Rev.\ Lett.\ }{\bf #1},}
\def\MPLA#1,{{\jnfont Mod.\ Phys.\ Lett.\ A }{\bf #1},}
\def\JPG#1,{{\jnfont J.\ Phys.\ G }{\bf #1},}
\def\CTP#1,{{\jnfont Commun.\ Theor.\ Phys.\ }{\bf #1},}
\def\JHEP#1,{{\jnfont JHEP \ }{\bf #1},}
\def\NPPS#1,{{\jnfont Nucl.\ Phys.\ Proc.\ Suppl.\ }{\bf #1},}
\def\CPC#1,{{\jnfont Computl.\ Phys.\ Commun.\ }{\bf #1},}
\def\CPL#1,{{\jnfont Chin.\ Phys.\ Lett. }{\bf #1},}
\def\AJS#1,{{\jnfont Astrophys.\ J.\ Suppl. }{\bf #1},}
\def\PR#1,{{\jnfont Phys.\ Rept. }{\bf #1},}
\def\AP#1,{{\jnfont Astropart.\ Phys. }{\bf #1},}
\def\EPL#1,{{\jnfont Europhys.\ Lett. }{\bf #1},}
\def\FP#1,{{\jnfont Fortsch.\ Phys. }{\bf #1},}

\begin{document}

\title{\ \\[10mm] Dark matter and Higgs phenomenology
                  predicted by left-right twin Higgs model
                  in light of CDMS II results
                  }

\author{Lei Wang$^1$, Jin Min Yang$^{2}$}

\affiliation{
$^1$ Department of Physics, Yantai University, Yantai 264005, PR China\\
$^2$ Key Laboratory of Frontiers in Theoretical Physics,\\
     Institute of Theoretical Physics, Academia Sinica,
             Beijing 100190, PR China}


\begin{abstract}
The left-right twin Higgs model predicts a light stable scalar $\hat{S}$,
which is a candidate for WIMP dark matter.
We study its scattering on nucleon and find
that the cross section is below the CDMS II upper bound but can
reach the SuperCDMS sensitivity. Then we study the Higgs
phenomenology by paying special attention to the decay $h\to
\hat{S}\hat{S}$ which is strongly correlated with the dark matter
scattering on nucleon. We find that such an invisible decay can be
sizable, which can severely suppress the conventional decay modes
like $h\to VV (V=W,~Z)$ and $h\to b\bar{b}$. On the other hand,
compared to the SM prediction, the rates of Higgs boson productions
at the LHC via gluon-gluon fusion, weak boson fusion or in
association with top quark pairs are all reduced significantly,
e.g., the gluon-gluon fusion channel can be suppressed by about
$30\%$.
\end{abstract}


\maketitle

\section{Introduction}
The twin Higgs mechanism \cite{twinhiggs,lrth} is proposed as
an interesting solution to the hierarchy problem. The SM Higgs
emerges as a pseudo-Goldstone boson once a global symmetry is
spontaneously broken, which is similar to what happens in the little
Higgs models \cite{littlehiggs}. An additional discrete symmetry is
imposed, which ensures the absence of one-loop quadratic divergence
of Higgs mass. The resulting Higgs boson mass is naturally around the
electroweak scale when the cut-off scale of the theory is around 5-10
TeV. The twin Higgs mechanism can be implemented in left-right
models with the additional discrete symmetry being identified as
the left-right symmetry \cite{lrth}. In the left-right twin Higgs
(LRTH) model, several physical Higgs bosons remain after the spontaneous
symmetry breaking. Another additional discrete symmetry is
introduced in the model under which the $SU(2)_L$ doublet $\hat{h}$
is odd while all the other fields are even. The lightest particle
$\hat{S}$ in its neutral components is stable and thus can be a
candidate for weakly interacting massive particle (WIMP) dark
matter. The phenomenology of LRTH model has been studied by some
authors \cite{otherwork}.

The density of cold dark matter in the universe has been determined
precisely by WMAP \cite{wmap}:
\beq
\Omega_{CDM}h^2=0.105^{+0.021}_{-0.030}.
\eeq
The thermal production of WIMPs can naturally explain such a relic density.
As a direct detection of WIMPs, the CDMS attempts to
observe the recoil energy transferred to a target nucleus in an
elastic collision with a WIMP. Very recently the CDMS collaboration
has completed their analysis of the final data runs of the CDMS II
experiment and reported two candidate events \cite{cdms2}. Although
these events cannot be interpreted as significant evidence for WIMP
interacting with nucleons, the CDMS gives the most stringent upper
limit on the WIMP-nucleon spin-independent cross section. For
example, the cross section is constrained to be smaller than
$3.8\times10^{-44}$ cm$^2$ for a WIMP of 70 GeV at 90\% confidence
level \cite{cdms2}. The implications of the new results from the
CDMS II experiment have been discussed in many models \cite{papercdms}.

In this work we focus on the left-right twin Higgs model. We first
examine the scattering of the dark matter candidate $\hat{S}$ with
nucleon and compare the rate with the CDMS II results. Then we
study the Higgs phenomenology, paying special attention to the
decay $h\to \hat{S}\hat{S}$ which is strongly correlated with the
dark matter scattering on nucleon. We will figure out the size of
such an invisible decay rate and how severely to suppress the
conventional decay modes like $h\to VV (V=W,~Z)$ and $h\to
b\bar{b}$. We also study the suppression for the rates of Higgs
boson productions at the LHC via gluon-gluon fusion, weak boson
fusion or in association with a pair of top quarks. Since the LHC
will be able to discover the Higgs boson in the full mass range
\cite{lhchiggs}, our study will help to probe the left-right
twin Higgs model.

This work is organized as follows. In Sec. II, we briefly review the
left-right twin Higgs model. In Sec. III, we examine the scattering of
the dark matter candidate $\hat{S}$ with nucleon and compare the rate
with the CDMS II results. Also, the correlation of Higgs decays with
the dark matter scattering on nucleon is studied.
In Sec. IV, we calculate the main productions of the Higgs boson
at the LHC. Finally, we give our conclusion in Sec. V.

\section{left-right twin Higgs model}

\subsection{Mass terms of gauge bosons}
In LRTH model \cite{lrth,phlrth}, the global symmetry is $U(4)\times
U(4)$ with a gauged $SU(2)_L\times SU(2)_R\times U(1)_{B-L}$
subgroup. The twin symmetry is identified as a left-right symmetry
which interchanges $L$ and $R$, implying that that gauge couplings of
$SU(2)_L$ and $SU(2)_R$ are identical ($g_{2L}=g_{2R}=g_2$).

A pair of Higgs fields, $H$ and $\hat{H}$, are introduced and each
transforms as $(\textbf{4},\textbf{1})$ and
$(\textbf{1},\textbf{4})$ respectively under the global symmetry.
They can be written as
\begin{equation}
H=\left(
\begin{tabular}{c}
$H_L$\\
$H_R$
\end{tabular}
\right),\ \ \ \ \ \hat{H}=\left(
\begin{tabular}{c}
$\hat{H}_L$\\
$\hat{H}_R$
\end{tabular}
\right),
\end{equation}
where $H_{L,R}$ and $\hat{H}_{L,R}$ are two component objects which
are charged under ${\rm SU}(2)_L\times {\rm SU}(2)_R\times {\rm
U}(1)_{B-L}$ as
\begin{equation}
    H_L {\rm \ and\ }\hat{H}_L: ({\bf 2},{\bf 1},1),\ \ \
    H_R {\rm \ and\ }\hat{H}_R: ({\bf 1},{\bf 2},1).
\end{equation}
Each Higgs acquires a non-zero VEV as
\begin{eqnarray}\label{eq:vev1}
    <H> = \left(%
\begin{array}{c}
  0 \\
  0 \\
  0 \\
  f \\
\end{array}%
\right),\;\;\;\;\;
<\hat{H}> = \left(%
\begin{array}{c}
  0 \\
  0 \\
  0 \\
  \hat{f} \\
\end{array}%
\right),
\end{eqnarray}
which breaks one of the U(4) to U(3) and yields seven
Nambu-Goldstone bosons. The scalar fields can be parameterized as
\begin{equation}
    H = f e^{i\frac{\pi}{f}}\left(%
\begin{array}{c}
  0  \\
  0  \\
  0  \\
  1  \\
\end{array}%
\right),\ \ \ \ \ {\rm with\ }
    \pi = \left(%
\begin{array}{cccc}
  -N/2 & 0 & 0 & h_1 \\
  0 & -N/2 & 0 & h_2 \\
  0 & 0 &  -N/2 & C \\
  h_1^* & h_2^* & C^* & 3N/2  \\
\end{array}%
\right), \label{eq:Higgsrep}
\end{equation}
with $\pi$ being the corresponding Goldstone fields. $N$ is a neutral
real pseudoscalar, $C$ and $C^*$ are a pair of charged complex
scalar fields, and $(h_1,h_2)^T$ is the SM ${\rm SU}(2)_L$ Higgs
doublet. $\hat{H}$ can be parameterized in the same way by its own
Goldstone fields $\hat{\pi}$, which contains $\hat{N}$, $\hat{C}$
and $\hat{h}=(\hat{h}_1^+, \hat{h}_2^0)^T$.

The generators of $SU(2)_L\times SU(2)_R\times U(1)_{B-L}$ are
given respectively as
\begin{equation}
\left(\begin{array}{cc} \frac{1}{2}\sigma_i & 0 \\ 0 &
0\end{array}\right),\qquad \left(\begin{array}{cc} 0 & 0 \\ 0 &
\frac{1}{2}\sigma_i\end{array}\right),\qquad
\frac{1}{2}\left(\begin{array}{cc} 1_2 & 0 \\ 0 & 1_2
\end{array}\right),
\end{equation}
and the corresponding gauge fields are
\begin{equation}
    W_2 =\frac{1}{2}\left(%
\begin{array}{cccc}
  W^0_L & \sqrt{2}W^+_L & 0 & 0 \\
  \sqrt{2}W^-_L & -W^0_L & 0 & 0 \\
   0 & 0 & W^0_R & \sqrt{2}W^+_R \\
   0 & 0 & \sqrt{2}W^-_R & -W^0_R \\
\end{array}%
\right),\ \ \
    W_{B-L} =\frac{W_1}{2}
\left(%
\begin{array}{cccc}
  1 & 0& 0 & 0 \\
 0 & 1 & 0 & 0 \\
   0 & 0 & 1& 0 \\
   0 & 0 & 0 & 1 \\
\end{array}%
\right),
\end{equation}
where the Lorentz indices are suppressed. The covariant derivative is
\begin{equation}
 D^{\mu} = \partial^{\mu} - i g_2 W_2^{\mu} -ig_1n_{B-L}W_{B-L}^{\mu},
\end{equation}
where $g_1$ and $g_2$ are the gauge couplings for ${\rm U}(1)_{B-L}$
and ${\rm SU}(2)_{L,R}$, and $n_{B-L}$ is the charge of the field
under ${\rm U}(1)_{B-L}$.

The covariant kinetic terms of Higgs fields can be written down as
\cite{lrth,phlrth}
\begin{eqnarray}
    {\cal L}_H &=& (D_{\mu}H)^{\dagger}D^{\mu}H +
(D_{\mu}\hat{H})^{\dagger}D^{\mu}\hat{H},
\end{eqnarray}
with $n_{B-L}=1$. The above Lagrangian contains the following
neutral Higgs boson interactions: \bea {\cal L}_H &\supset &
\frac{1}{2} g_2^2 f^2 s_1^2 W^-_L W^+_L + \frac{1}{2} g_2^2
(\hat{f}^2 + f^2 c_1^2) W^-_R W^+_R + \frac{1}{4} g_1^2 (f^2 +
\hat{f}^2) W_1 W_1 \nonumber\\&-& \frac{1}{4} g_1 g_2 f^2(1-c_{2})
 W_1 W^0_L + \frac{1}{8} g_2^2 f^2(1-c_2)
 W^{0}_L W^{0}_L - \frac{1}{4} g_1 g_2 (f^2 + f^2 c_2 + 2\hat{f}^2) W_1 W^0_R\nonumber \\&+& \frac{1}{8} g_2^2
 (f^2 + f^2 c_2 + 2\hat{f}^2) W^{0}_R W^{0}_R,\eea
where\begin{eqnarray}
    \label{yi}
    \begin{array}{rclrcl}
    c_1&=&\cos\frac{h+v}{\sqrt{2}f},
                   \qquad &
    s_1&=&\sqrt{1-c_1^2}, \\
    c_2&=&\cos\frac{\sqrt{2}(h+v)}{f}, \qquad &
    s_2&=&\sqrt{1-c_2^2}.
    \end{array}
\end{eqnarray}
The $h$ and $v$ are the SM-like Higgs field and its VEV,
respectively, which arise from the $SU(2)_L$ doublet $(h_1,h_2)^T$.
For the charged gauge bosons, there is no mixing between $W_L^\pm$
and $W_R^\pm$: $W^{\pm}=W_L^{\pm}$ and $W_H^{\pm}=W_R^{\pm}$. At
$\ord{(\frac{v^2}{f^2})}$, their masses and Higgs couplings are
\begin{eqnarray}
    \label{yi}
    \begin{array}{rclrcl}
     m^2_{W}&=&\frac{1}{4}g_2^2 v^2 (1-\frac{v^2}{6f^2}),
                   \qquad &
     m^2_{W_H}&=&\frac{1}{2}g_2^2 [\hat{f}^2+f^2(1-\frac{v^2}{2f^2})], \\
    hWW&:&\frac{1}{2}g_2^2 v(1-\frac{v^2}{3f^2}), \qquad &
    hW_H W_H&:&-\frac{1}{2}g_2^2 v(1-\frac{v^2}{3f^2}).
    \end{array}
\end{eqnarray}
The neutral gauge bosons $Z_H$, $Z$ and $\gamma$ are linear
combinations of $W_L^0$, $W_R^0$ and $W_1$. Ref. \cite{phlrth} gives
the leading-order masses and Higgs couplings for the  mass eigenstates.
The diagonalization of the gauge mass matrix is performed
numerically in our analysis, and the coupling of $hZZ$ can be
obtained at $\ord{(\frac{v^2}{f^2})}$.

\subsection{Mass terms of fermions}
The masses of the first two generation quarks and bottom quark are
obtained from the non-renormalizable operators \cite{phlrth}
\begin{equation}
    {\cal L}_{Y}=\frac{y_u^{\alpha\beta}}
{\Lambda}(\bar{Q}_{L\alpha}\tau_2 H_L^*)(H_R^T\tau_2{Q}_{R\beta})
+\frac{y_d^{\alpha\beta}}{\Lambda}(\bar{Q}_{L\alpha}
H_L)(H_R^{\dagger}{Q}_{R\beta}) + h.c.,
\label{Yukawa2}
\end{equation}
where $\tau_2=\left(\begin{array}{cc} 0 & -1 \\ 1 & 0
\end{array}\right)$, $Q_{L\alpha}=-i(u_{L\alpha},d_{L\alpha})^T$ and
$Q_{R\alpha}=(u_{R\alpha},d_{R\alpha})^T$ with $\alpha$ being the
family index. For simplicity, we assume the quark flavor mixing is
small and neglect the mixing effects. From Eq. (\ref{Yukawa2}), we
can get the Higgs boson interactions with the first two generation
quarks and bottom quark:
\begin{equation}
   {\cal L}_{Y}\simeq -\frac{y_u^{\alpha}}
{2\Lambda}f^2 s_2 \bar{u}_{L\alpha}u_{R\alpha} -\frac{y_d^{\alpha}}
{2\Lambda}f^2 s_2\bar{d}_{L\alpha}d_{R\alpha} + h.c.
\label{Yukawamass}
\end{equation}
The mass and Higgs coupling of the quark $q$ are given by
\beq
m_q= \frac{y_q}{\sqrt{2}} \frac{f} {\Lambda}v (1-\frac{v^2}{3f^2}),\ \ \
\
h\bar{q}q:-\frac{m_q}{v}(1-\frac{2}{3}\frac{v^2}{f^2}),
\label{bcoupling}
\eeq
where $q$ denotes the first two generation quarks or bottom quark.

For the lepton sector, the Yukawa interaction is similar to Eq.
(\ref{Yukawa2}), which can generate small masses for the charged
leptons and the Dirac mass terms for neutrinos.

For the top quark Yukawa interaction, in order to cancel the
one-loop quadratic divergence of Higgs mass induced by the top
quark, a pair of vector-like quarks $(U_L,U_R)$ are introduced.
The Lagrangian can be written as \cite{phlrth}
\begin{equation}
{\cal L}_{t}=y_L\bar{Q}_{L3}\tau_2 H_L^*U_R
  +y_R\bar{Q}_{R3}\tau_2H_R^*U_L - M\bar{U}_LU_R + h.c.
\label{yukawatop}
\end{equation}
where $Q_{L3} = -i(u_{L3},d_{L3})^T$ and $Q_{R3} =
(u_{R3},d_{R3})^T$. Under left-right symmetry, $y_L=y_R=y$. From
Eq.(\ref{yukawatop}), we can get Higgs interaction as \beq {\cal
L}_{t}\simeq -y f s_1 \bar{u}_{L3} U_R - y f c_1\bar{u}_{R3} U_L - M
\bar{U}_L U_R +h.c. \label{topmass} \eeq By diagonalizing the mass
matrix in Eq. (\ref{topmass}), we obtain the mass eigenstates for
the top quark and heavy top quark partner $T$. The field $t_L$ and
$T_L$ ($t_R$ and $T_R$) are the linear combination of $u_{L3}$ and
$U_L$ ($u_{R3}$ and $U_R$), respectively. The masses and Higgs
couplings of the mass eigenstates are given by \cite{phlrth}
\begin{eqnarray}
    \label{yi}
    \begin{array}{rclrcl}
     m^2_t&=&\frac{1}{2}(M^2+y^2f^2-N_t),
                   \qquad &
     m^2_T&=&\frac{1}{2}(M^2+y^2f^2+N_t), \\
    h\bar{t}t&:&-\frac{m_t}{v} C_L C_R, \qquad &
    h\bar{T}T&:&-\frac{y}{\sqrt{2}} (S_R S_L-C_L C_R x).
    \end{array}
\end{eqnarray}
where
\begin{eqnarray}
    \label{yi}
    \begin{array}{rclrcl}
    S_L&=&\frac{1}{\sqrt{2}}\sqrt{1-(y^2f^2\cos2x+M^2)/N_t},
                   \qquad &
    C_L&=&\sqrt{1-S_L^2}, \\
    S_R&=&\frac{1}{\sqrt{2}}\sqrt{1-(y^2f^2\cos2x-M^2)/N_t}, \qquad &
    C_R&=&\sqrt{1-S_R^2},\\
    N_t&=&\sqrt{(y^2f^2+M^2)^2-y^4f^4\sin^22x},
    \end{array}
\end{eqnarray}
with $x=\frac{v}{\sqrt{2}f}$.

\subsection{Mass term of dark matter}
In addition to the Coleman-Weinberg potential arising from gauge
boson contributions, the soft left-right symmetry breaking terms, so
called ``$\mu$-term", can give masses for $\hat{h}_1^\pm$ and
$\hat{h}_2^0$ \cite{phlrth}:
\begin{equation}
V_{\mu}=-\mu^2_r(H^\dagger_R\hat H_R+h.c.)+\hat \mu^2 \hat
H^\dagger_L \hat H_L. \label{muterm}
\end{equation}
In order not to reintroduce fine tuning, $\mu_r$ should be less than
about $f/4\pi$. It is natural for $\hat \mu$ not to be much larger
than $f$. The masses of $\hat{h}_2^0$ and $\hat{h}_1^\pm$ are
\begin{eqnarray}
M^2_{\hat h_2}&=&\frac{3}{16\pi^2}\Big[\frac{g_2^2}{2} ({\mathcal
Z}(M_W)-{\mathcal Z}(M_{W_H}))
+\frac{2g_1^2+g_2^2}{4}\frac{M^2_{W_H}-M^2_W}{M^2_{Z_H}-M^2_Z}
({\mathcal Z}(M_Z)-{\mathcal Z}(M_{Z_H}))\Big]\nonumber\\
&&\,+\mu^2_r\frac{f}{\hat f}\cos x+\hat \mu^2,\nonumber\\
M^2_{\hat h_1}&\simeq&M^2_{\hat h_2}, \label{h2h1mass}
\end{eqnarray}
where ${\mathcal Z}(x)=-x^2(\ln\frac{\Lambda^2}{x^2}+1)$, and the
cut-off scale $\Lambda$ is typically taken to be $4 \pi f$. We
neglect the small mass splitting between $\hat{h}_2^0$ and
$\hat{h}_1^\pm$ due to the electromagnetic interactions. Note that
$\hat{\mu}^2$ could have either sign, which can allow us to vary the
masses of $\hat{h}_2^0$ and $\hat{h}_1^\pm$ as a free parameter.

The complex scalar $\hat{h}_2^0$ can be written as
\begin{equation}
\hat{h}_2^0=\frac{\hat{S}+i\hat{A}}{\sqrt{2}},
\end{equation}
where $\hat{S}$ and $\hat{A}$ are the scalar and pseudoscalar
fields, respectively. We can introduce a new quartic potential term
to get the mass splitting between $\hat{S}$ and $\hat{A}$, as well
as their Higgs couplings \cite{lrthdm}:
\begin{equation}
V_H=-\frac{\lambda_5}{2} [(H_L^\dagger\hat{H}_L)^2+h.c.].
\label{lam5}
\end{equation}
From Eq. (\ref{lam5}), we can get
\begin{eqnarray}
    \label{lam5has}
    \begin{array}{rclrcl}
    \delta
    m_{\hat{S}}^2&=&-\frac{\lambda_5}{2}v^2(1-\frac{v^2}{6f^2}),
                   \qquad &
   \delta m_{\hat{A}}^2&=&\frac{\lambda_5}{2}v^2(1-\frac{v^2}{6f^2}), \\
    h\hat{S}\hat{S}&:&\lambda_5 v (1-\frac{v^2}{3f^2}), \qquad &
    h\hat{A}\hat{A}&:&-\lambda_5 v (1-\frac{v^2}{3f^2}).\\
     \end{array}
\end{eqnarray}
Since the quartic terms $-(\lambda_5/4) h^2 \hat{S}^2$ and
$(\lambda_5/4) h^2 \hat{A}^2$ induced by Eq. (\ref{lam5}) have
opposite sign, the one-loop quadratic divergence of Higgs mass
from the $\hat{S}$ loop and from the $\hat{A}$ loop can be
cancelled. Therefore, it is safe to take $\lambda_5 \sim 1$.

There is also a quartic term which can potentially introduce a mass
splitting between $\hat{h}_2^0$ and $\hat{h}_1^\pm$ \cite{lrthdm}:
\begin{equation}
V'_H=\lambda_4 |\hat{H}_L^\dagger H_L|^2. \label{lam4}
\end{equation}
However, unlike Eq. (\ref{lam5}), Eq. (\ref{lam4}) can produce a
dangerous contribution to the Higgs mass if $\lambda_4$ is too
large, which requires $| \lambda_4 |\leq \frac{1}{16\pi^2}$ with
$\Lambda=4\pi f$. Therefore, compared with Eq. (\ref{lam5has}), the
corrections of Eq. (\ref{lam4}) to the $\hat{h}_2^0$ mass and Higgs
coupling can be neglected. We define two parameters:
\begin{eqnarray}
\delta_2 &\equiv&
m_{\hat{A}}-m_{\hat{S}}=\frac{\lambda_5}{(m_{\hat{A}}
       +m_{\hat{S}})}v^2(1-\frac{v^2}{6f^2}),\nonumber\\
\delta_1&\equiv& m_{\hat{h}_1}-m_{\hat{S}}=\frac{\lambda_5}{2(m_{\hat{h}_1}
  +m_{\hat{S}})}v^2(1-\frac{v^2}{6f^2}).
\label{del21}
\end{eqnarray}
From Eqs. (\ref{h2h1mass}), (\ref{lam5has}) and (\ref{del21}), we
can get the relation $\delta_2\approx 2\delta_1$ when $m_{\hat{S}}$
is much larger than $\delta_2$. Actually, we checked that
for a value of $m_{\hat{S}}$ not much larger than $\delta_2$,
the relation $\delta_2\approx 2\delta_1$ is still a good approximation.
For example, for ($m_{\hat{S}},~\delta_2$)=(34~GeV,~24~GeV),
(70~GeV,~40~GeV) and (70~GeV,~20~GeV), we found that
$\delta_2/\delta_1$ is 1.77, 1.80 and 1.88, respectively.
In our numerical calculations we assume $\delta_2 = 2\delta_1$
and we checked that the results are changed very little if we take
$\delta_2/\delta_1$ differently as  1.77, 1.80 or 1.88.

$\hat{S}$ is lighter than $\hat{A}$, and can be a candidate of dark
matter. In addition to the Higgs couplings in Eq. (\ref{lam5has}),
the Coleman-Weinberg potential can give the contributions to the
couplings of $h\hat{S}\hat{S}$, $h\hat{A}\hat{A}$ and
$h\hat{h}_1\hat{h}_1$. These expressions are complicated and can be
found in \cite{suintel}. In our analysis, these contributions are
considered.

\section{Dark matter scattering on nucleon and Higgs decay}
In LRTH model, the neutral $\hat{S}$ is a candidate for WIMP dark
matter. Ref. \cite{lrthdm} shows that there are two distinctive mass
regions for $\hat{S}$ which can give a relic density in the WMAP $3
\sigma$ range: (i) low mass region, and (ii) high mass region. In
this paper we focus on the low mass region where the invisible decay
$h\to \hat{S}\hat{S}$ can be open, which can change other decay
branching ratios and thus affect the strategy of searching for the
Higgs boson at high energy colliders. For such a low mass region of
$\hat{S}$, Fig. 9 in Ref. \cite{lrthdm} shows that in the region
satisfying the constraints of $\Gamma_Z$ and WMAP $3 \sigma$ relic
density, $\delta_2$ can vary in the range of 20 GeV and 40 GeV for
$m_{\hat{S}}\approx70$ GeV, and for 30 GeV $\lesssim
m_{\hat{S}}\lesssim70$ GeV the value of $\delta_2$ is around 20 GeV.

In LRTH model, the elastic scattering of $\hat{S}$ on a nucleus
receives the dominant contributions from the Higgs boson exchange
diagrams. The spin-independent cross section between $\hat{S}$ and
the nucleon is given by \cite{sigmasi}
\beq
\sigma_{\hat{S}
p(n)}^{SI}=\frac{m_{p(n)}^{2}}{4\pi\left(m_{\hat{S}}+m_{p(n)}\right)^{2}}
    \left[f^{p(n)}\right]^{2},
\eeq where \beq
f^{p(n)}=\sum_{q=u,d,s}f_{T_{q}}^{p(n)}\mathcal{C}_{\hat{S}
q}\frac{m_{p(n)}}{m_{q}}+\frac{2}{27}f_{T_{g}}^{p(n)}\sum_{q=c,b,t}\mathcal{C}_{\hat{S}
q}\frac{m_{p(n)}}{m_{q}},\label{fpn} \eeq with \cite{fpfn}
\begin{eqnarray}
f_{T_{u}}^{(p)}\approx0.020,\quad & f_{T_{d}}^{(p)}\approx0.026,\quad & f_{T_{s}}^{(p)}\approx0.118,\quad  f_{T_{g}}^{(p)}\approx0.836,\nonumber \\
f_{T_{u}}^{(n)}\approx0.014,\quad &
f_{T_{d}}^{(n)}\approx0.036,\quad & f_{T_{s}}^{(n)}\approx0.118,
\quad f_{T_{g}}^{(n)}\approx0.832,\nonumber\\
 \mathcal{C}_{\hat{S} q}=\frac{g_{h\hat{S}\hat{S}}g_{h\bar{q}q}}{m_h^2}.
\label{eq:neuclon-form}
\end{eqnarray}
Here $g_{h\hat{S}\hat{S}}$ and $g_{h\bar{q}q}$ are the couplings of
$h\hat{S}\hat{S}$ and $h\bar{q}q$. Note that $\sigma_{\hat{S}
p}^{SI}\approx \sigma_{\hat{S} n}^{SI}$.

In this model the major decay modes of the Higgs boson are the
SM-like ones: $h\to f \bar{f}$ (SM fermion pair), $WW$ and $ZZ$.
The LRTH model gives corrections to these decay modes via the
corresponding modified Higgs couplings
\beq
\Gamma(h \to XX)= \Gamma(h \to
XX)_{SM}(g_{hXX}/g_{hXX}^{SM})^2,
\end{equation}
where $XX$ denotes fermion pairs, $WW$ or $ZZ$. $\Gamma(h \to
XX)_{SM}$ is the decay width in the SM, and $g_{hXX}$ and
$g_{hXX}^{SM}$ are the couplings of $hXX$ in the LRTH model and SM,
respectively. In our calculations the relevant higher order QCD and
electroweak corrections are considered using the code Hdecay
\cite{hdecay}. In addition to the SM-like decay modes, the Higgs
boson has some new important decay modes which are kinematically
allowed in some parameter space: $h\to\hat{S}\hat{S}$, $h\to
\hat{A}\hat{A}$ and $h \to \hat{h}_1\hat{h}_1$, whose partial widths
are given by \bea \label{eq:Gamma:new} \Gm(h \to \hat{S}\hat{S}) &=&
 \frac{g_{h\hat{S}\hat{S}}^2}{32\pi m_h}\sqrt{1-x_{\hat{S}}},\nonumber\\
\Gm(h \to \hat{A}\hat{A}) &=&
 \frac{g_{h\hat{A}\hat{A}}^2}{32\pi m_h}\sqrt{1-x_{\hat{A}}},\nonumber\\
\Gm(h \to \hat{h}_1\hat{h}_1) &=&
 \frac{g_{h\hat{h}_1\hat{h}_1}^2}{16\pi m_h}\sqrt{1-x_{\hat{h}_1}},
\eea
where $x_s =4m_s^2/m_h^2$ with $s=\hat{S},~\hat{A}$ and $\hat{h}_1$.

In our calculations, the free parameters involved are $f$,
$\Lambda$, $M$, $\mu_r$, $m_{\hat{S}}$ and $\delta_2$. $\hat{f}$ can
be determined by $f$, $\Lambda$, $M$ and $\mu_r$ by requiring that
the SM Higgs obtains an electroweak symmetry breaking VEV of 246 GeV
\cite{phlrth}. The Higgs mass depends on $f$, $\Lambda$, $M$ and
$\mu_r$. Note that in our numerical calculations we used the exact
expressions (not only keeping the leading terms of $v^2/f^2$).
The Z-pole precision measurements, low energy neutral
current process and high energy precision measurements off the
Z-pole can give strong constraints on $f$, requiring approximately
$f>500$ GeV.
Of course, the fine tuning becomes severe if $f$ is too large
\cite{phlrth}. Following Ref. \cite{suintel}, we take the typical
parameter space: $500$ GeV $\leq f\leq$ $1500$ GeV, $\Lambda=4\pi
f$, $\mu_r=50$ GeV and $M=0~(150)$ GeV, where the Higgs mass is
approximately in the range of $160$ GeV and $180$ GeV.

\begin{figure}[tb]
 \epsfig{file=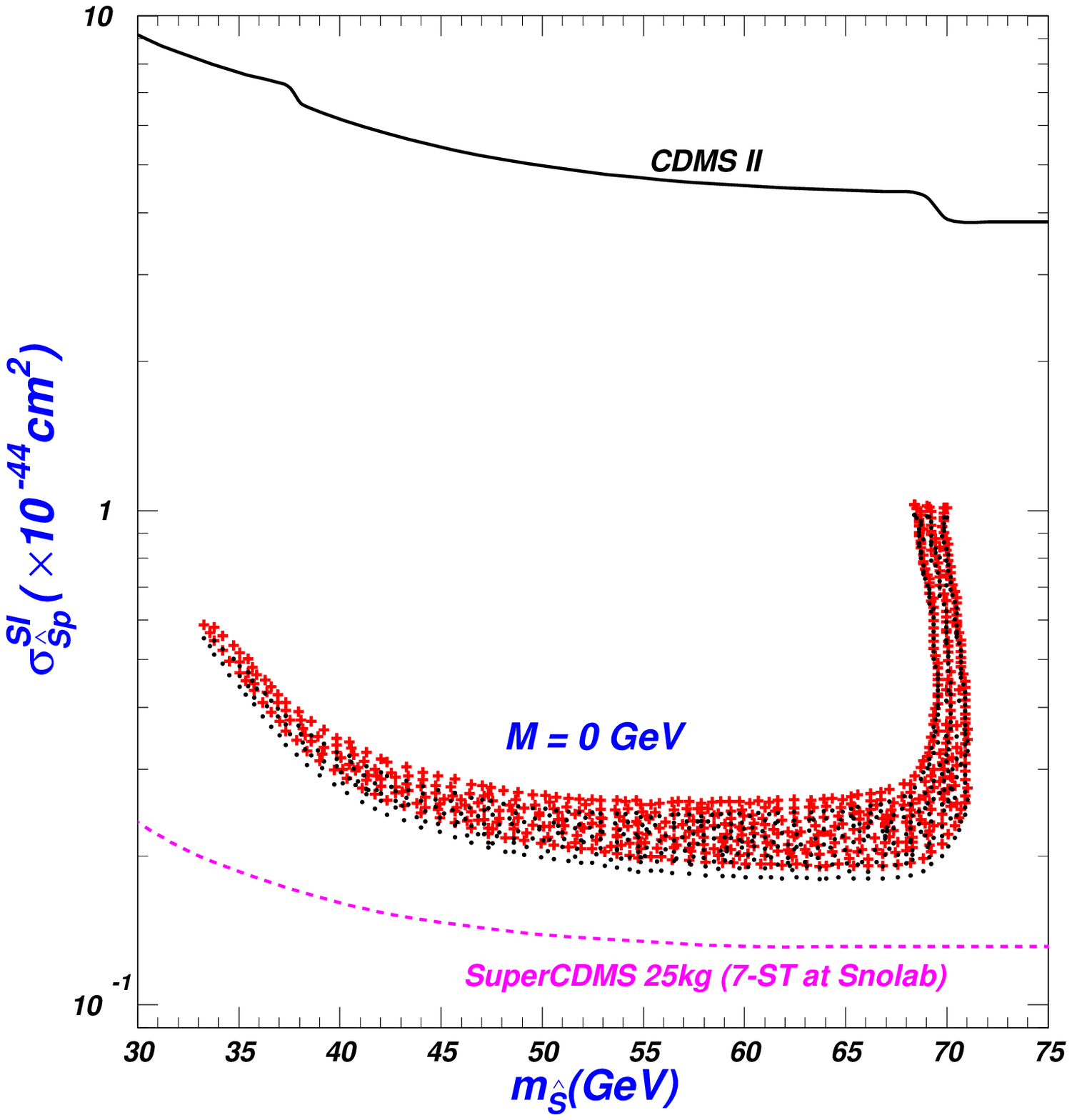,height=7.5cm}
 \epsfig{file=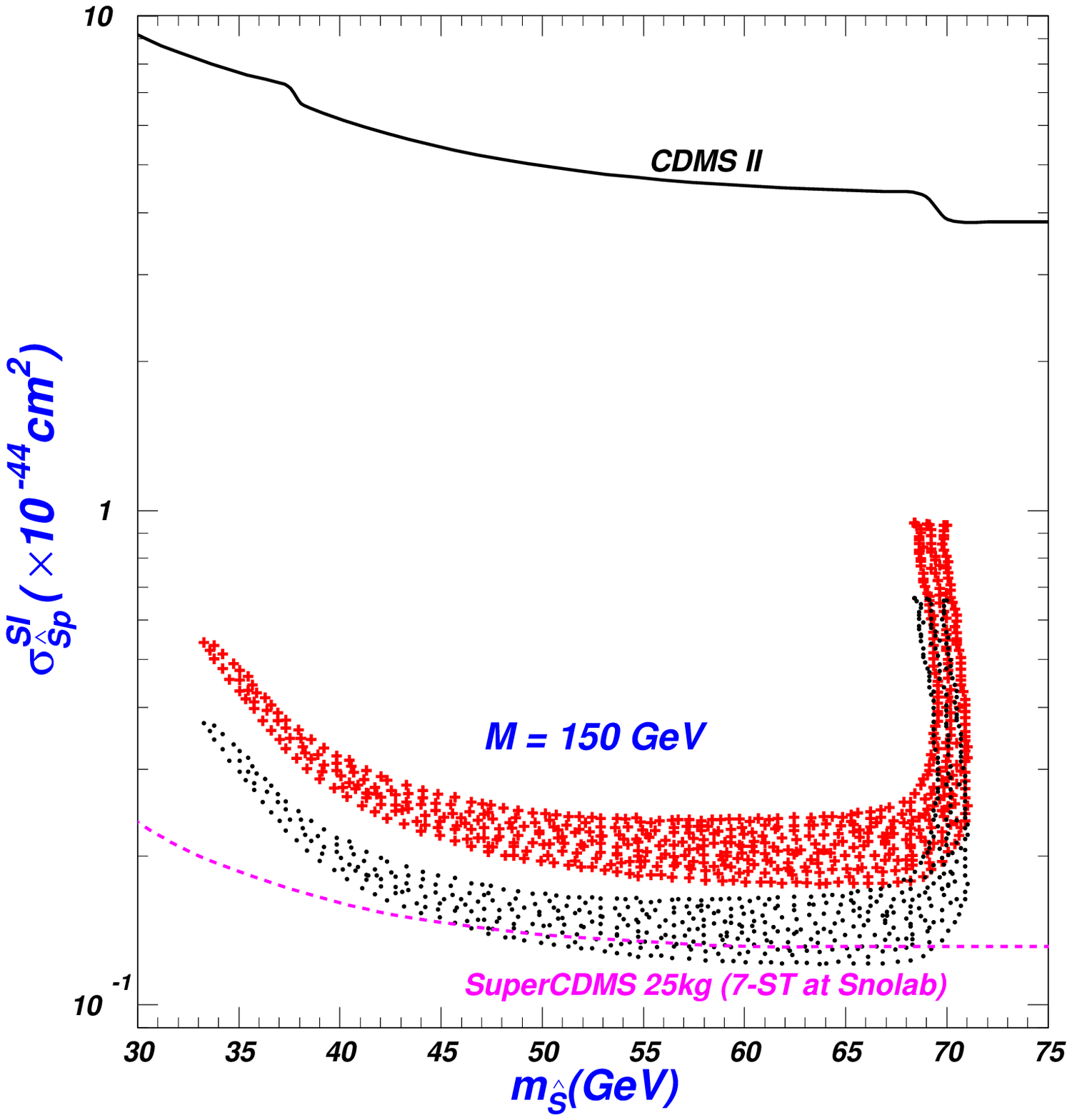,height=7.5cm}
\vspace{-.5cm} \caption{Scatter plots of $\sigma_{\hat{S} p}^{SI}$
versus $m_{\hat{S}}$ from the scan over the parameters $m_{\hat{S}}$
and $\delta_2$ in the region satisfying the constraints of
$\Gamma_Z$ and WMAP $3 \sigma$ relic density \cite{lrthdm}. The
lower region denoted by bullets (black) and the upper region denoted
by crosses (red) are for $f=500$ GeV and $f=1$ TeV, respectively. }
\label{cdms}
\end{figure}
\begin{figure}[tb]
 \epsfig{file=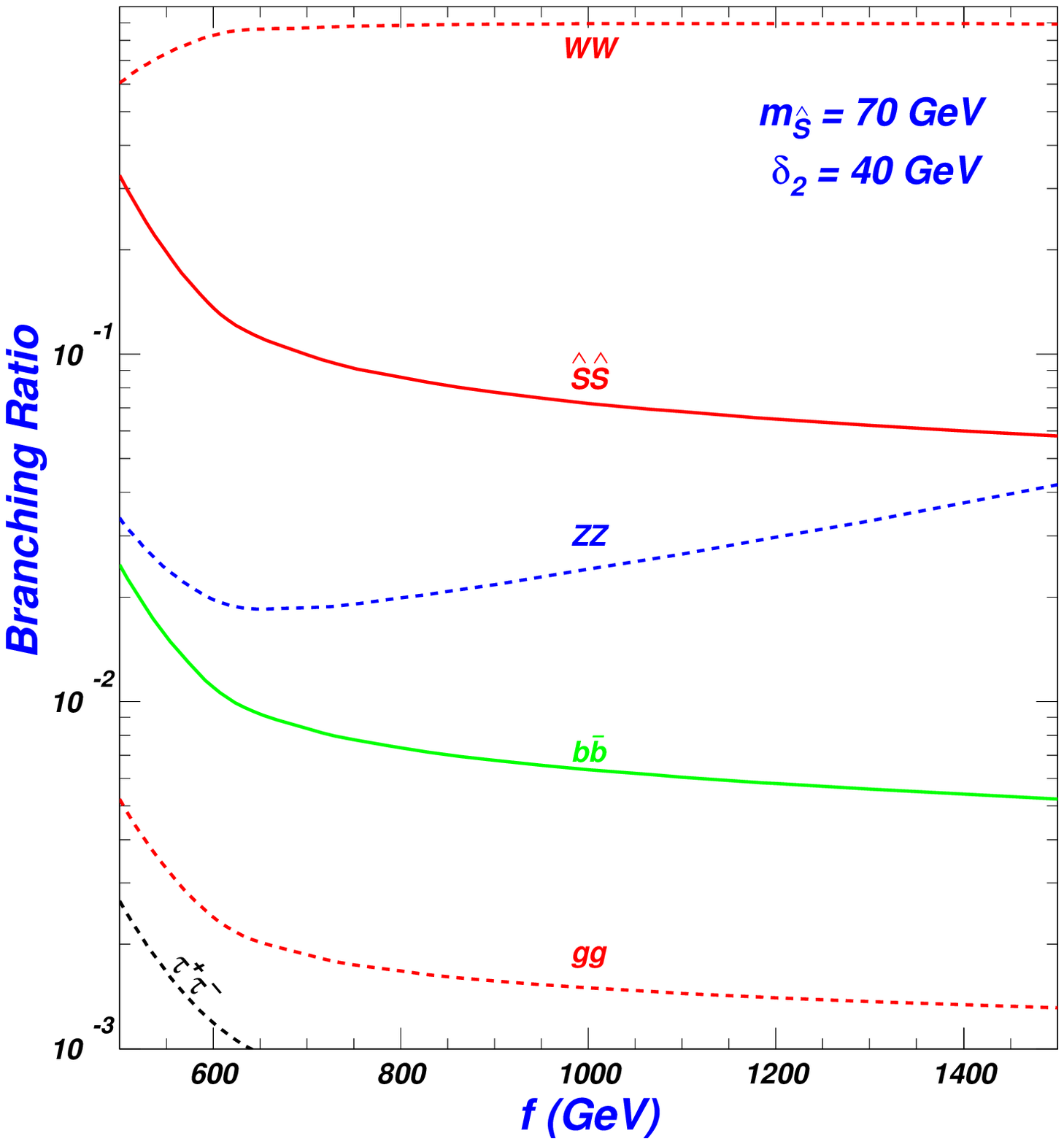,height=7.5cm}
 \epsfig{file=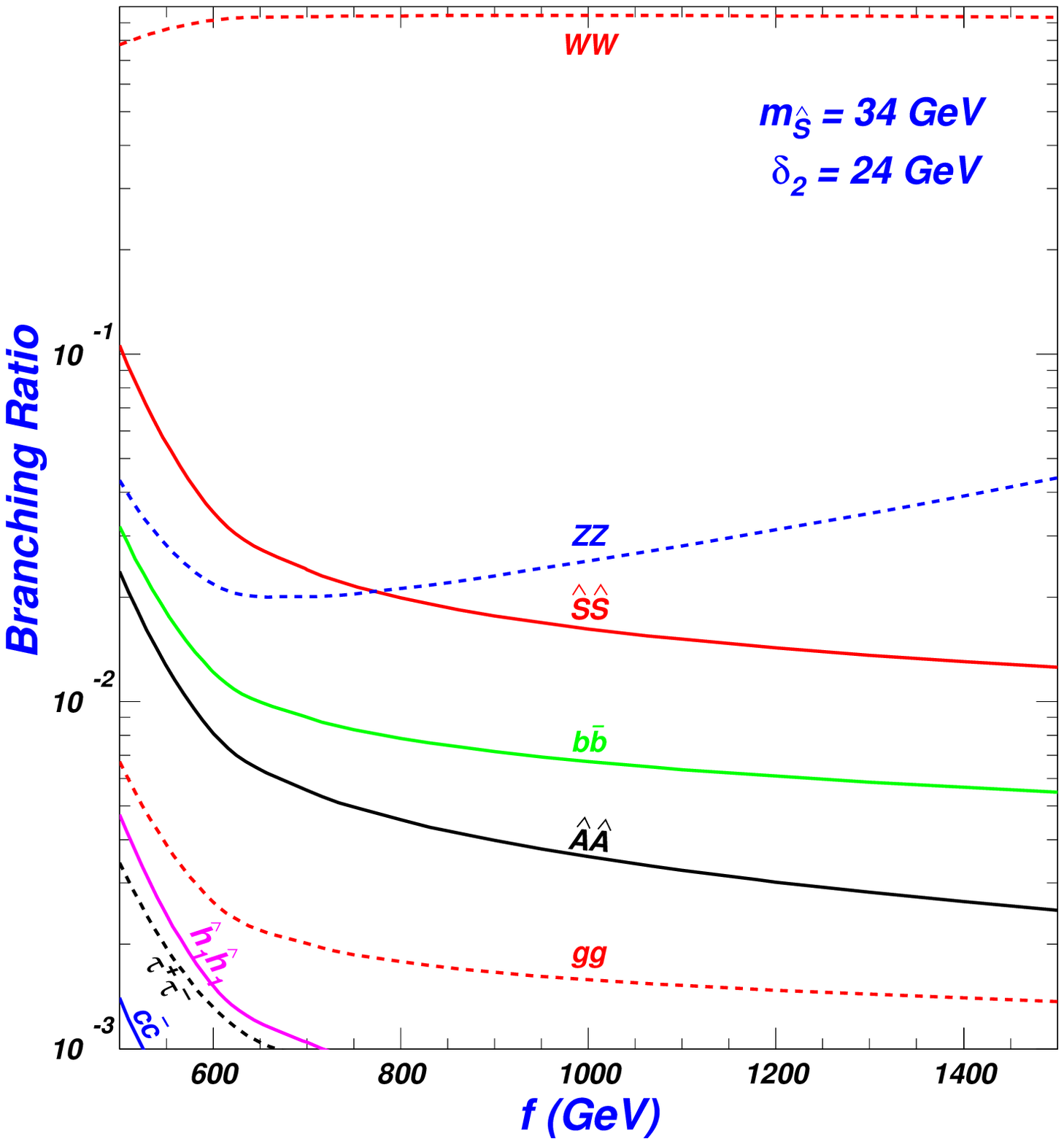,height=7.5cm}
\vspace{-.5cm} \caption{The Higgs decay branching ratios versus the
scale $f$ for $M=0$ GeV.} \label{br}
\end{figure}

We scan over $m_{\hat{S}}$ and $\delta_2$ in the region satisfying
the constraints of WMAP $3 \sigma$ relic density and $\Gamma_Z$. The
scatter plots of $\sigma_{\hat{S} p}^{SI}$ versus $m_{\hat{S}}$ are
displayed in Fig. \ref{cdms}. We see that $\sigma_{\hat{S} p}^{SI}$
is well below the CDMS II upper bound for the low mass region of
$\hat{S}$. The cross section can reach $1.0\times10^{-44}$ cm$^2$
for $m_{\hat{S}}\approx70$ GeV where $\delta_2=40$ GeV is allowed.
The coupling of $h\hat{S}\hat{S}$ increases with $\delta_2$, which
can enhance $\sigma_{\hat{S} p}^{SI}$ sizably. The parameters $M$
and $f$ can have some effects on $\sigma_{\hat{S} p}^{SI}$ by
changing the Higgs mass which can suppress $\sigma_{\hat{S}
p}^{SI}$. Besides, the couplings of $h\hat{S}\hat{S}$ and
$hq\bar{q}$ increase with $f$ (see Eq. (\ref{bcoupling}) and Eq.
(\ref{lam5has})), which can give contributions to enhance
$\sigma_{\hat{S} p}^{SI}$.

In Fig. \ref{cdms} we also show the projected sensitivity of
SuperCDMS \cite{supercdms}. We see that the LRTH prediction
is accessible at SuperCDMS (25kg).

In Fig. \ref{br} we plot the Higgs decay branching ratios versus the
scale $f$ for $M=0$ GeV (note that the Higgs mass can be determined
by the value of $f$, e.g., $m_h$=159.3 GeV, 172.6 GeV and 178.4 GeV
for $f$=500 GeV, 1 TeV, 1.5 TeV, respectively). The left panel shows
that $Br(h\to \hat{S}\hat{S})$ is subdominant, and the largest value
can reach $32\%$ for $m_{\hat{S}}=70$ GeV, $\delta_2=40$ GeV and
$f=500$ GeV. The right panel shows that the new decay modes $h\to
\hat{A}\hat{A}$ and $h\to\hat{h}_1\hat{h}_1$ can be open for low
values of $m_{\hat{S}}$ and $\delta_2$, but their decay branching
ratios are relatively small. The branching ratios of all these three
new decay modes decrease sizably as $f$ increases. The reason is
that the Higgs mass increases with $f$, and the decay width of $h\to
WW$ becomes dominant for the large Higgs mass. The parameter $M$ can
have some effects on the Higgs decay modes mainly via changing the
Higgs mass, which are not shown here. Besides, $M$ can control the
couplings $ht\bar{t}$ and $hT\bar{T}$, which give the dominant
contributions to the decay $h\to gg$. We will show the dependence of
the decay $h\to gg$ on $M$ later.

\begin{figure}[htb]
 \epsfig{file=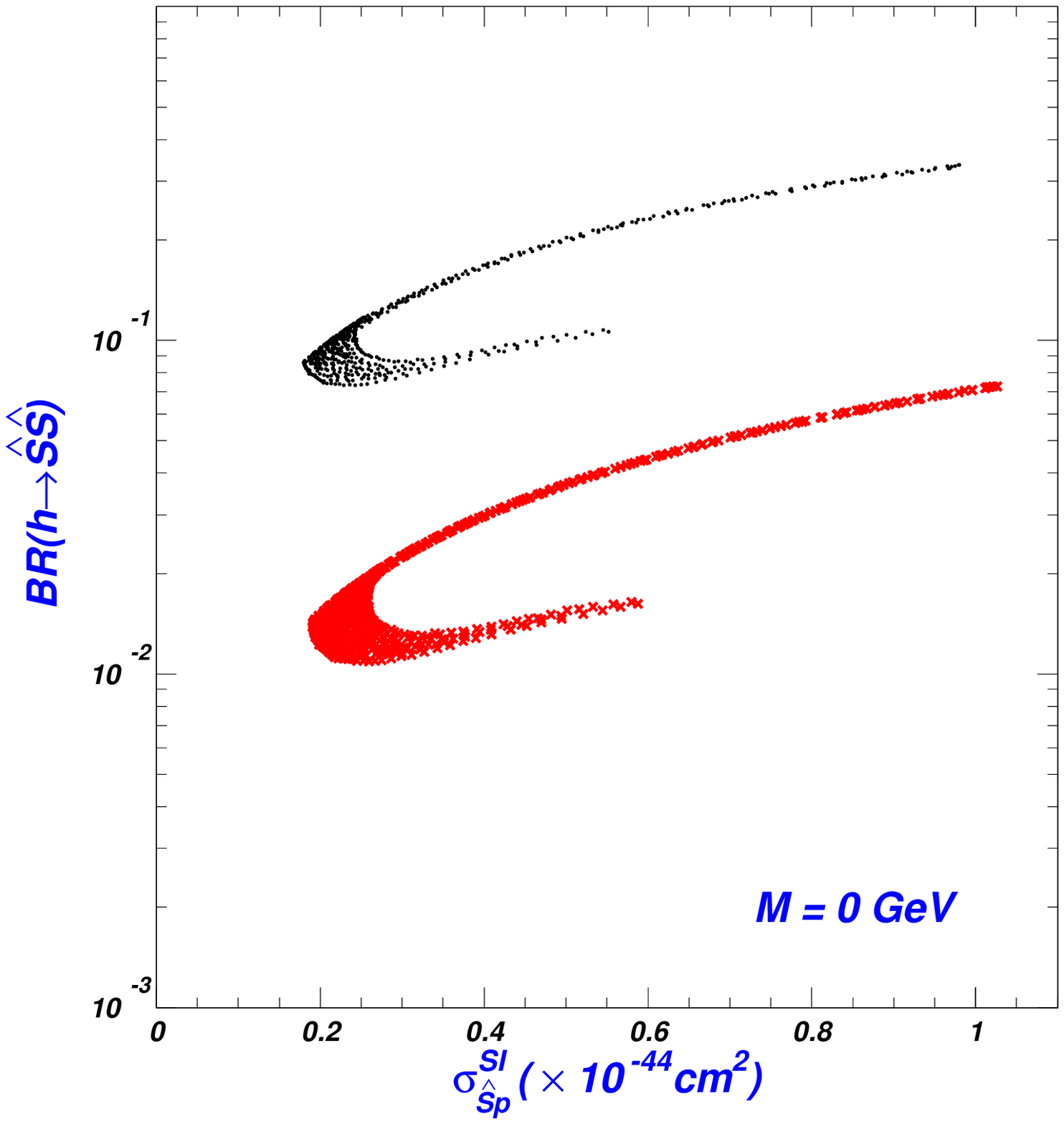,height=7.0cm}
 \epsfig{file=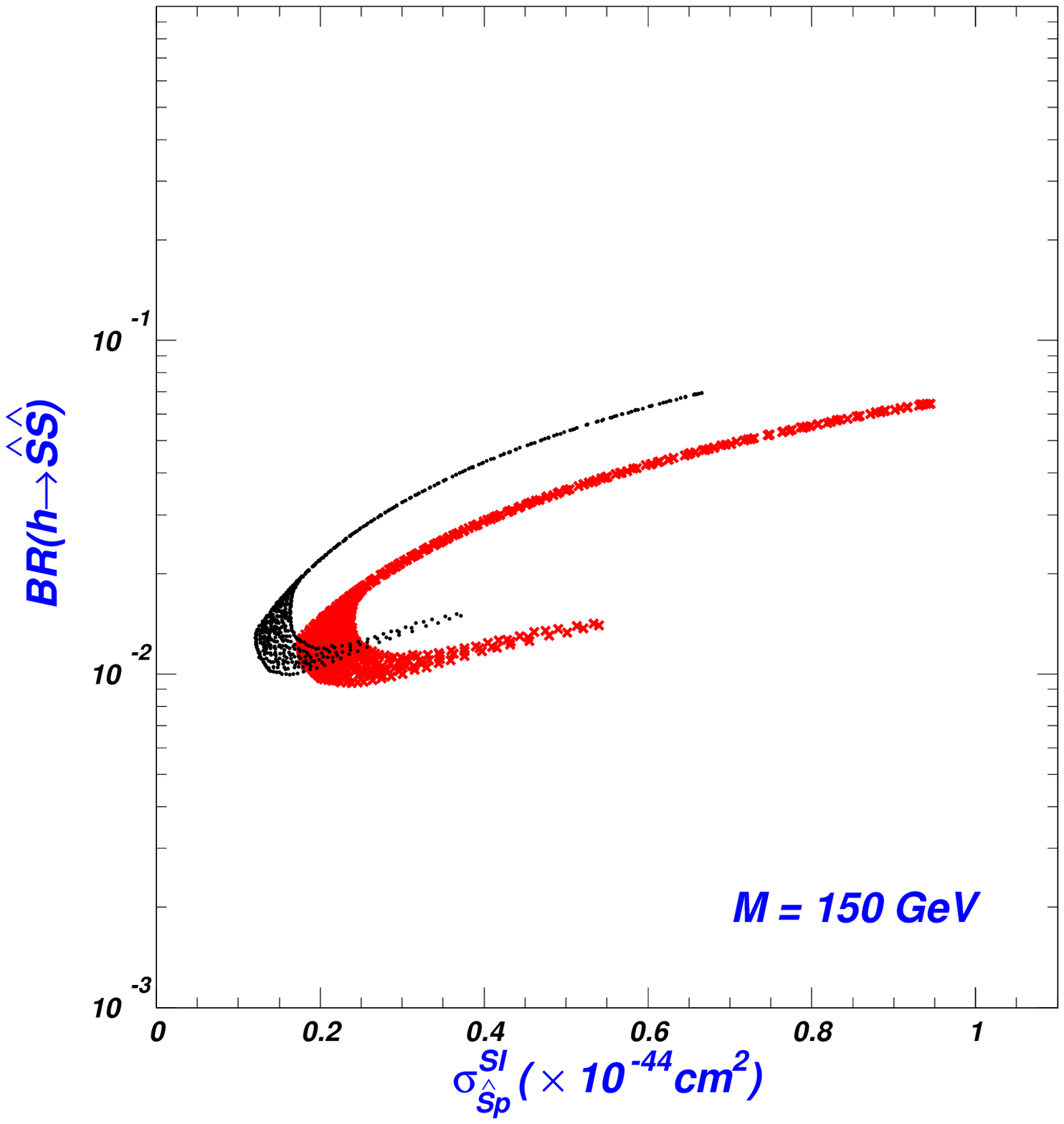,height=7.0cm}
\vspace{-.5cm} \caption{Same as Fig.\ref{cdms}, but projected on
the plane of $BR(h\to \hat{S}\hat{S})$ versus $\sigma_{\hat{S} p}^{SI}$.}
\label{brcdms}
\end{figure}

Fig. \ref{brcdms} shows the scatter plots of $BR(h\to
\hat{S}\hat{S})$ versus $\sigma_{\hat{S} p}^{SI}$ for $M=0$ GeV and
$M=150$ GeV, respectively. We can see that $BR(h\to \hat{S}\hat{S})$
is strongly correlated with $\sigma_{\hat{S} p}^{SI}$. When
$\sigma_{\hat{S} p}^{SI}$ increases, the corresponding $BR(h\to
\hat{S}\hat{S})$ also becomes large. So the SuperCDMS can probe
Higgs decay $h\to \hat{S}\hat{S}$ via measuring the spin-independent
WIMP-nucleon cross section, which is complementary to the
exploration of Higgs boson at high energy colliders.

\begin{figure}[htb]
 \epsfig{file=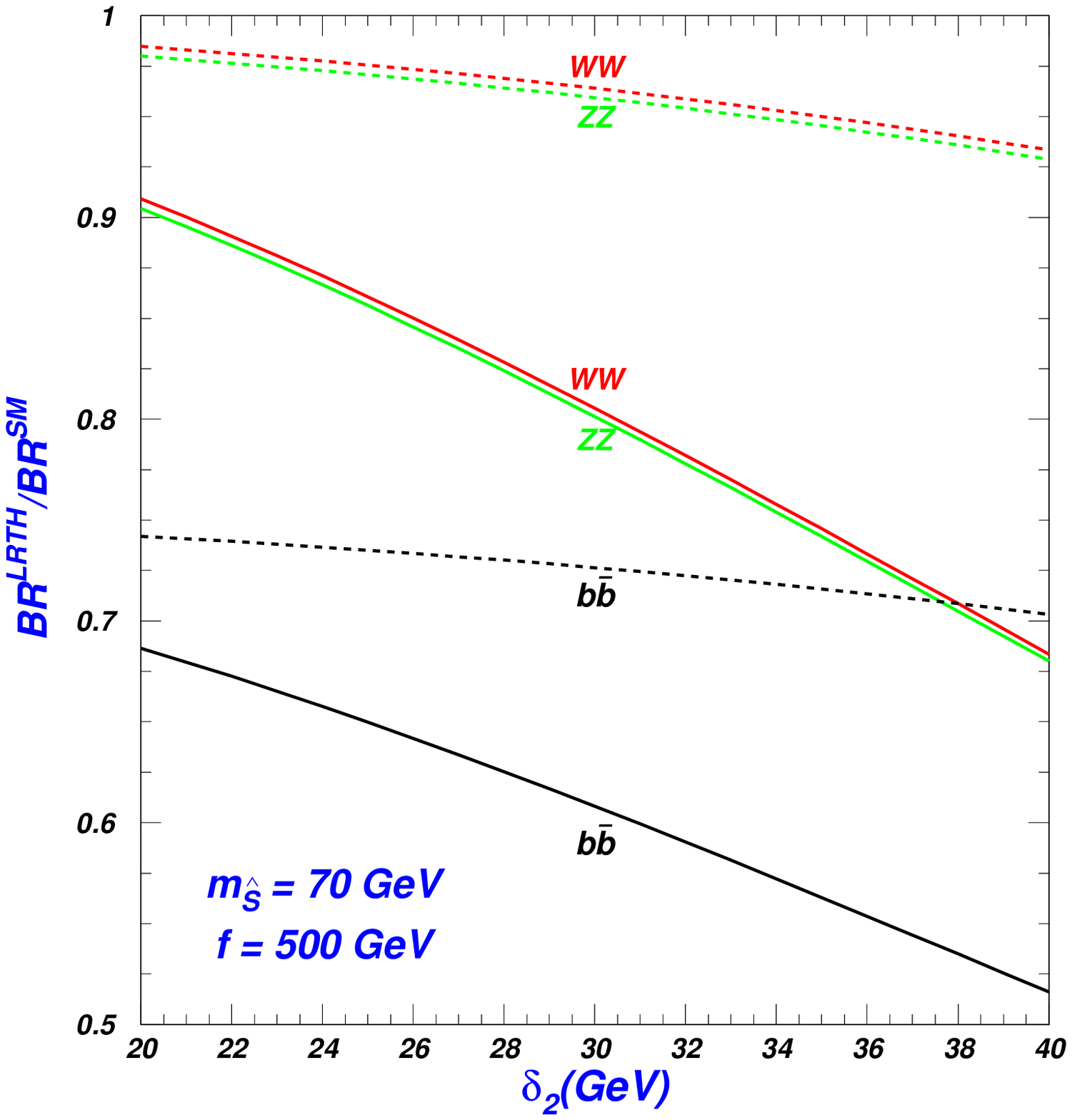,height=5.7cm}
 \epsfig{file=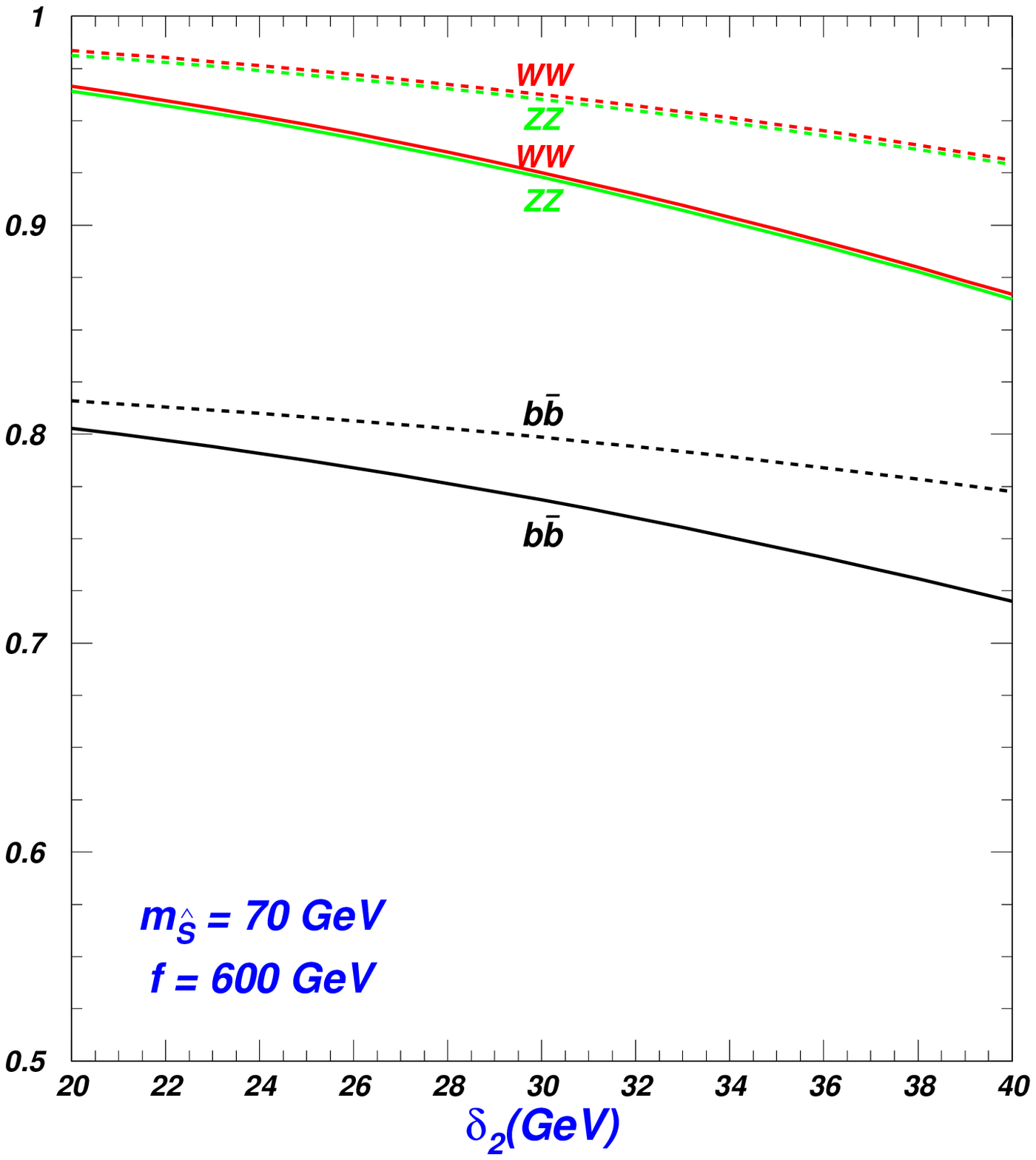,height=5.7cm}
 \epsfig{file=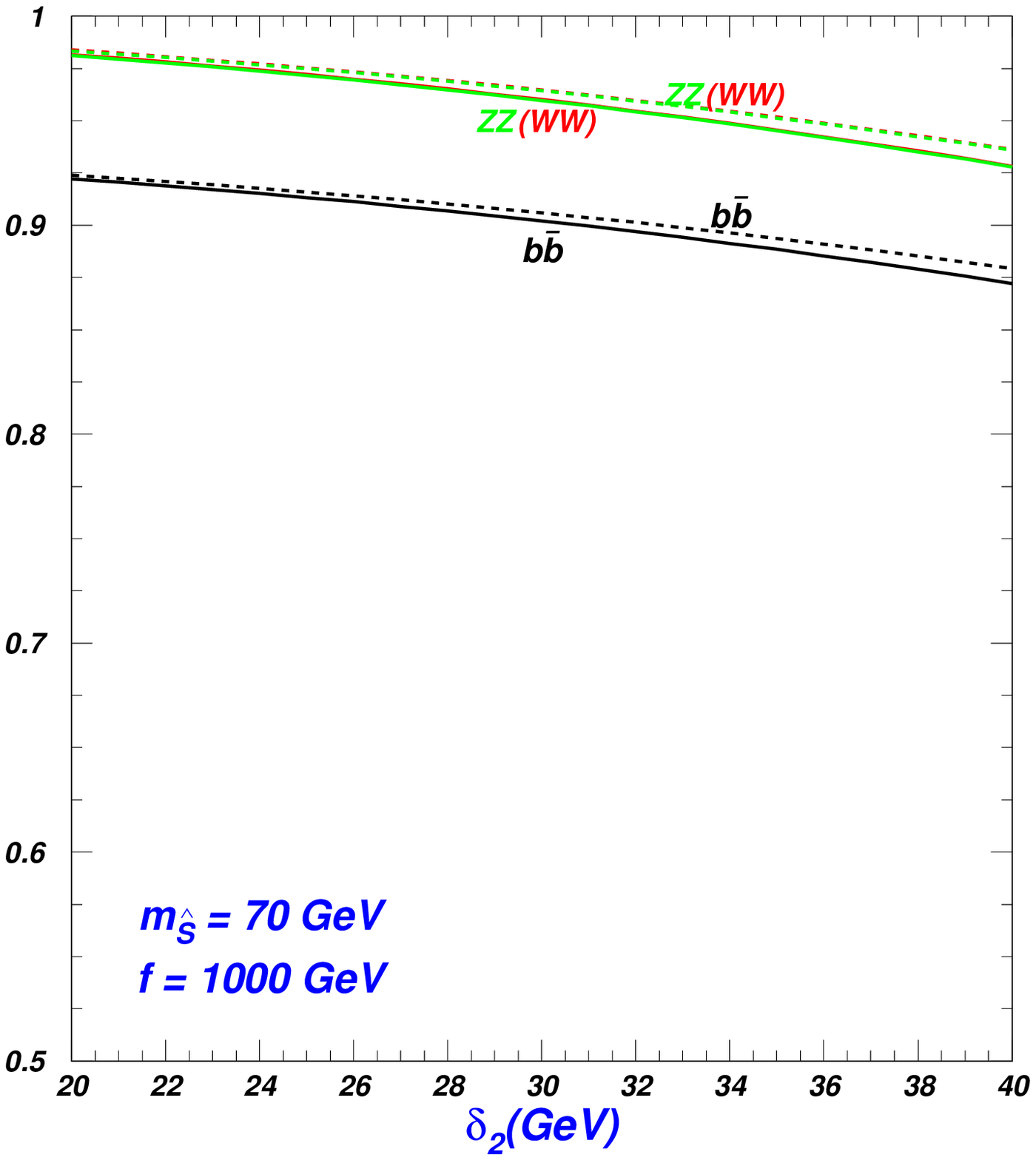,height=5.7cm}
 \epsfig{file=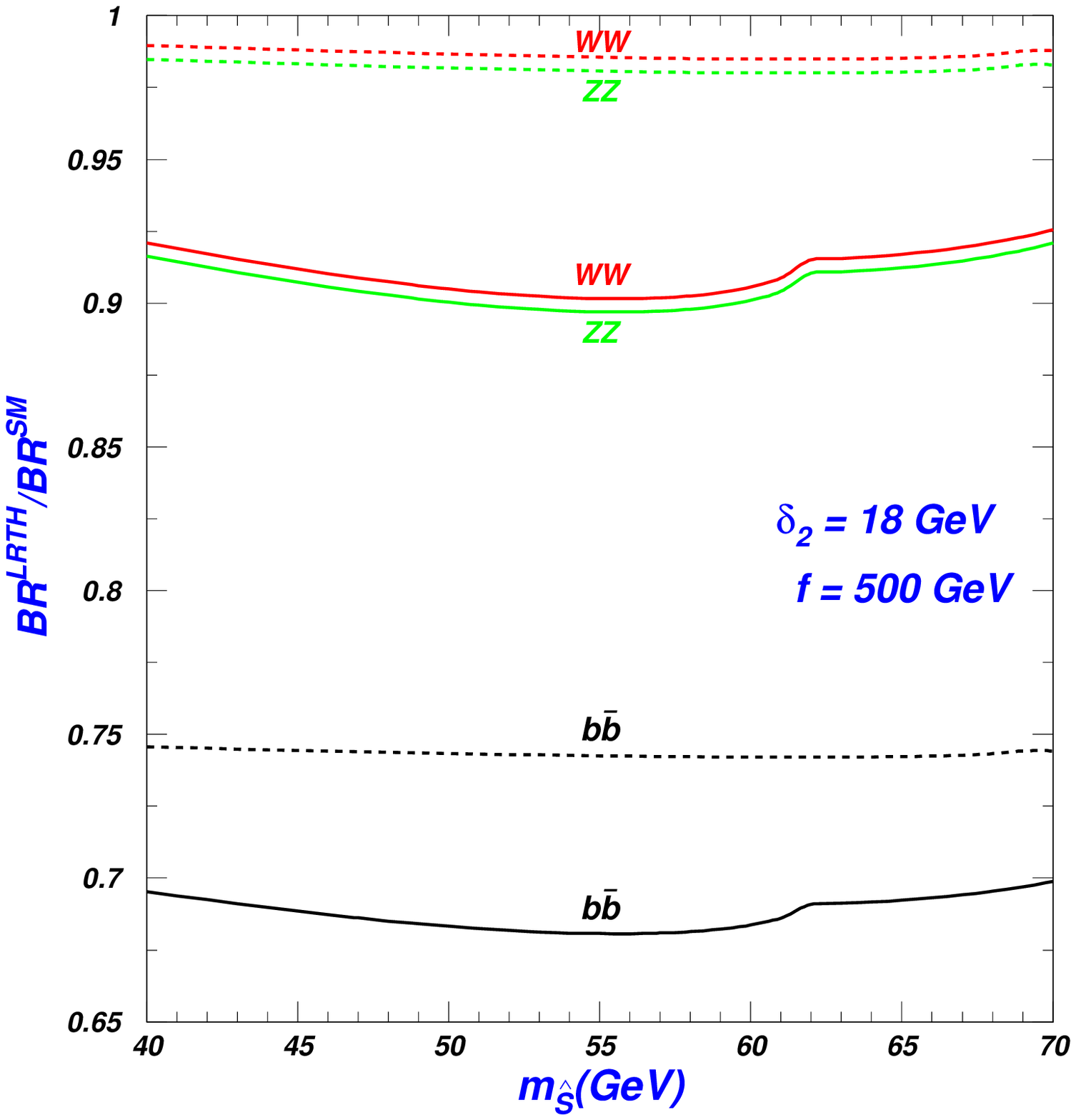,height=5.7cm}
 \epsfig{file=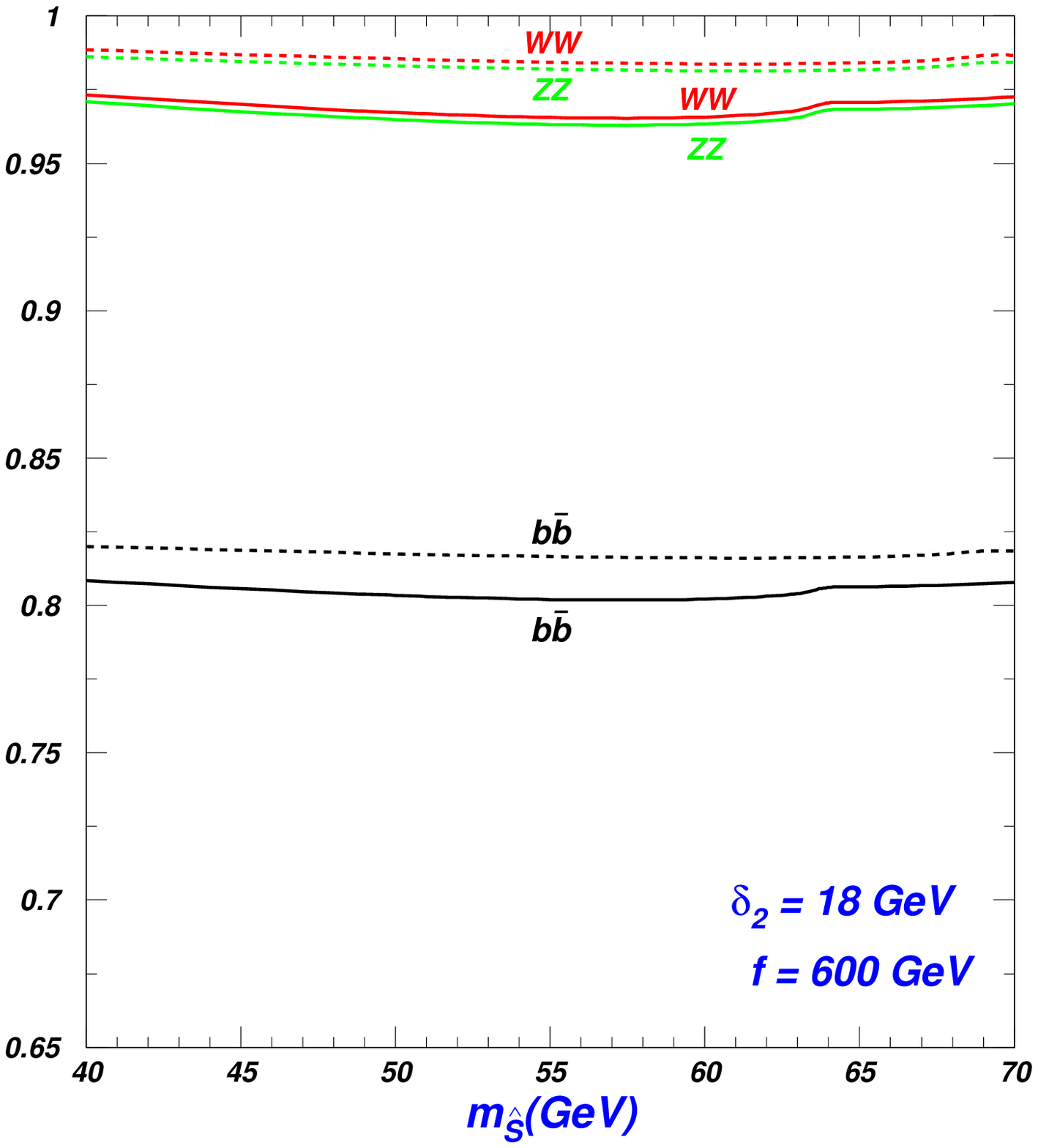,height=5.7cm}
 \epsfig{file=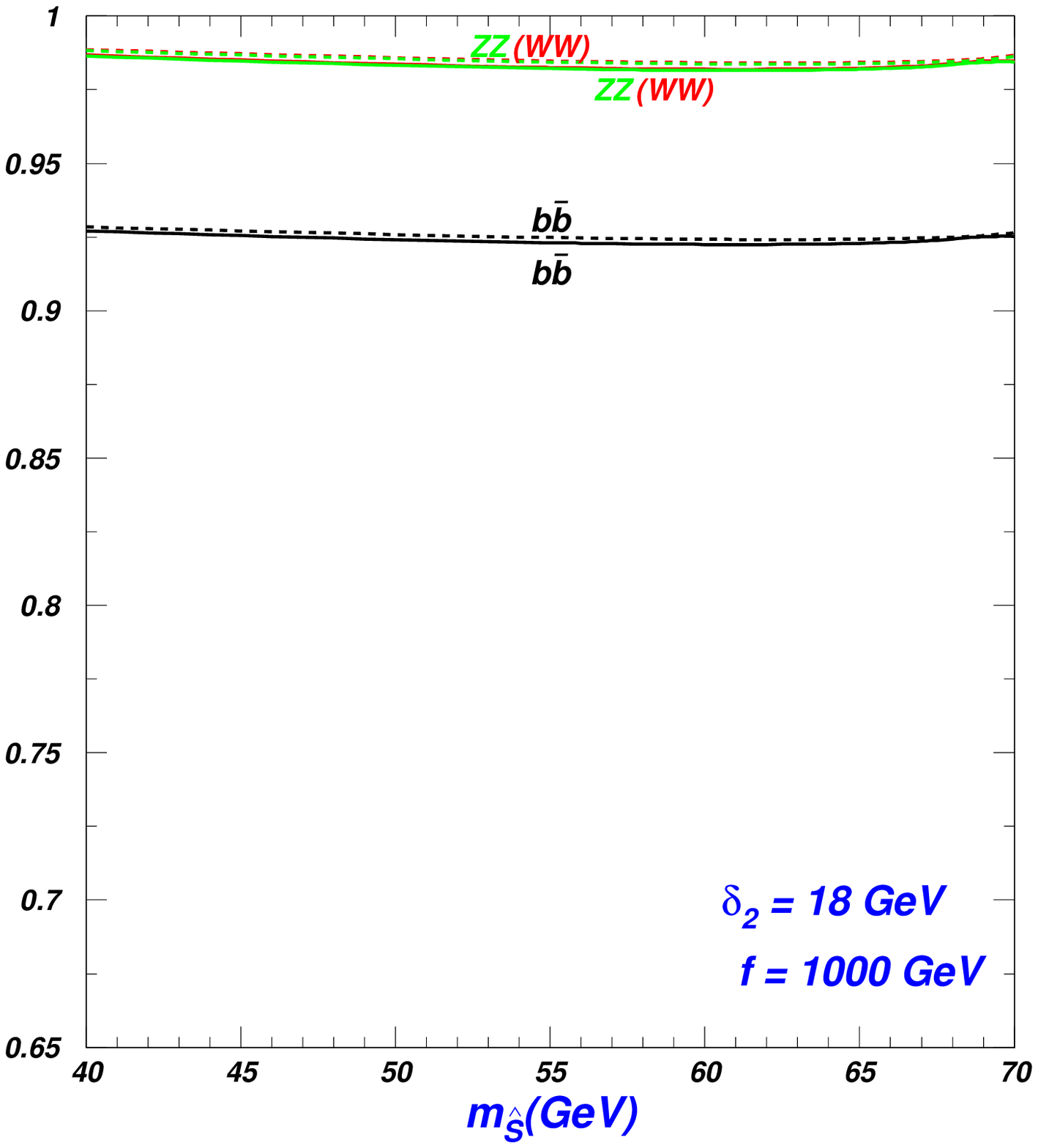,height=5.7cm}
\vspace{-.5cm} \caption{The decay branching ratios
$BR(h\to WW)$, $BR(h\to ZZ)$ and
$BR(h\to b\bar{b})$ normalized to the SM predictions. The
solid and dashed curves are for $M=0$ GeV and $M=150$ GeV,
respectively.} \label{brxiu}
\end{figure}

In Fig. \ref{brxiu} we plot $BR(h\to WW)$, $BR(h\to ZZ)$ and
$BR(h\to b\bar{b})$ normalized to the SM predictions for several
values of $f$. We see that the deviation from the SM prediction for
each decay mode is sensitive to $\delta_2$, and becomes more sizable
as $\delta_2$ increases. The corrections to $Br(h\to WW)$ and
$Br(h\to ZZ)$ are almost equal. For $M=0$ GeV, $f=500$ GeV,
$m_{\hat{S}}=70$ GeV and $\delta_2=40$ GeV, the deviations for the
decays $h\to VV$ ($V=W,~Z$) and $h\to b\bar{b}$ can be over 30\% and
47.5\%, respectively. The deviations from the SM predictions are
also sensitive to $M$ and $f$ which can change Higgs mass.

In LRTH model the Higgs mass is typically in the range of
$160-180$ GeV and $BR(h\to \gamma\gamma)$ is severely suppressed.
For $\Gamma(h\to\gamma\gamma)$, the $W$-boson contributions
dominate over the top quark contributions \cite{wloop}. The extra
fermions and bosons can give some relatively small corrections,
and their contributions tend to cancel each other. Therefore, the
modified coupling $hWW$ will give the dominant corrections to
$\Gamma(h\to\gamma\gamma)$, and the suppression of
$Br(h\to\gamma\gamma)$ approximately equals to that of $Br(h\to
WW)$, which happens in littlest Higgs model with T-parity
\cite{hrr}.

\section{Higgs production at LHC}
\begin{figure}[htb]
 \epsfig{file=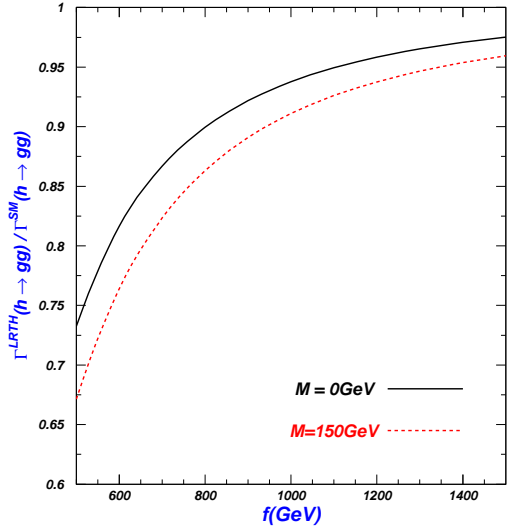,height=7.0cm}
\vspace{-.5cm} \caption{The ratio $\Gamma^{LRTH}(h\to
gg)/\Gamma^{SM}(h\to gg) =\sigma^{LRTH}(gg\to h)/\sigma^{SM}(gg\to
h)$ versus $f$.} \label{hgg}
\end{figure}

The Higgs production at the LHC is dominated by the gluon-gluon
fusion process. In the SM, the main contributions are from the top
quark loop, and the LRTH model can give corrections via the
modified coupling of $h\bar{t}t$ and the heavy T-quark loop. The
hadronic cross section $\sigma(gg\to h)$ has a
strong correlation with the decay width $\Gamma (h\to
 gg)$:
\bea
\sigma(gg\to h)&=&\hat{\sigma}(gg\to
h)\tau_0\int_{\tau_0}^1\frac{dx}{x}f_{g}(x,~\mu^2_F)f_{g}
(\frac{\tau_0}{x},~\mu^2_F),\nonumber\\
\hat{\sigma}(gg\to h)&=&\Gamma(h\to gg)\frac{\pi^2}{8m^3_h},
\label{ggtoh} \eea where $\tau_0=m^2_h/s$ with $\sqrt{s}$ being the
center-of-mass energy of the LHC. The parton distribution of gluon
$f_{g}$ is generated by CTEQ6L \cite{cteq6}, and both the
renormalization scale $\mu_R$ and the factorization scale $\mu_F$
are taken as $m_h$. From Eq. (\ref{ggtoh}) we get \beq
\frac{\sigma^{LRTH}(gg\to h)}{\sigma^{SM}(gg\to
h)}=\frac{\Gamma^{LRTH}(h\to gg)}{\Gamma^{SM}(h\to gg)}. \eeq The
width $\Gamma^{LRTH}(h\to gg)$ is independent of the parameters
$m_{\hat{S}}$ and $\delta_2$. In Fig. \ref{hgg}, we plot the ratio
$\Gamma^{LRTH}(h\to gg)/\Gamma^{SM}(h\to gg)$ versus the scale $f$
for $M=0$ GeV and $M=150$ GeV, respectively. We see that, compared
with the SM prediction, the LRTH model can suppress the partial
width sizably for a small value of $f$. As $f$ gets large, the
suppression is weakened. The deviation from the SM prediction is
also sensitive to the mixing parameter $M$. For $f=500$ GeV, the
suppression of SM prediction can reach 27\% and 32.5\% for $M=0$ GeV
and $M=150$ GeV, respectively.

Due to the suppression of the Higgs couplings with gauge bosons and
top quark, the Higgs boson production rates via the weak boson
fusion or in association with a pair of top quarks
are suppressed. Although $Br(h\to WW)$ can be
sizably suppressed in some parameter space, the decay mode $h\to WW$
is still an excellent channel for searching for Higgs boson with
an intermediate mass \cite{hwwhzz}.

In Table 1 the ratio $\frac{\sigma^{LRTH} \times
BR^{LRTH}}{\sigma^{SM}\times BR^{SM}}$ is listed for each main
channel: $gg\to h$, $VV\to h$ or $pp\to h\bar{t}t$ followed by $h\to
VV$ or $h\to b\bar{b}$. Table 1 shows that such a ratio of events
can be sizably suppressed in the LRTH model and the suppression
effects could exceed the experimental uncertainty (10\%-20\%)
\cite{uncertainty}. Therefore, it is possible to probe the LRTH
model via these Higgs production modes at the LHC.

Note that similar exotic decays for the SM-like Higgs boson may
also be predicted by some other new physics models like the little
Higgs models and supersymmetric or two
Higgs-doublet models \cite{leiwang}. A common feature of their phenomenology
is the suppression of the conventional visible channels of the Higgs
boson. To distinguish between different models,
all the channels of Higgs production should be jointly analyzed and
a linear collider is an ideal machine for such a purpose \cite{lei2}.

Recently both CDF and D0 Collaborations at the Tevatron have
searched for the SM Higgs boson and excluded the mass range
between 162 GeV and 166 GeV at 95\% CL \cite{tevatron}, using all
significant production modes: $gg\to h$, $VV\to h$, $q\bar{q}\to
Vh$ ($V=W,Z$) followed by $h\to WW$. The productions $gg\to h$ and
$VV\to h$ are suppressed, similar to the case at the LHC shown in
Table 1. The productions $q\bar{q}\to Vh$ are also suppressed,
just like $VV\to h$, because these productions are induced by the
suppressed couplings $hWW$ and $hZZ$. Since all these productions
$gg\to h$, $VV\to h$ and $q\bar{q}\to Vh$ followed by $h\to WW$
can be severely suppressed in the LRTH model, the constraints on
the Higgs mass from the Tevatron will be largely relaxed in the
LRTH model.

\begin{table}[t] \caption{The ratio $\frac{\sigma^{LRTH}\times
BR^{LRTH}}{\sigma^{SM}\times BR^{SM}}$ for various channels ( $gg\to
h$, $VV\to h$ or $pp\to h\bar{t}t$ followed by $h\to VV$ or $h\to
b\bar{b}$) with $m_{\hat{S}}=70$ GeV and $\delta_2=40$ GeV.}
 \vspace*{-0.8cm}
\begin{center}
\begin{tabular}{|ll|c|c|c|c|c|c|}
\hline  & & \multicolumn{3}{c|}{$h\to b \bar{b}$}
        &   \multicolumn{3}{c|}{$h\to VV$}\\
 \cline{3-8}
      & & $f$=500 GeV &  $f$=600 GeV &  $f$=1 TeV
        & $f$=500 GeV &  $f$=600 GeV &  $f$=1 TeV\\
\hline
$gg\to h$ &($M=0$ GeV) & 0.38& 0.59& 0.82
                      & 0.50& 0.71& 0.87\\
          &($M=150$ GeV)        & 0.47& 0.59& 0.80
                      &0.63 & 0.71& 0.85\\
\hline
$VV\to h$ &($M=0$ GeV)   &0.46 &0.66 &0.85 &0.60 &0.80 &0.90\\
          &($M=150$ GeV) &0.62 &0.71 &0.85 &0.82 &0.85 &0.91\\
\hline
$pp\to t\bar{t}h$ &($M=0$ GeV)&0.50 &0.70 &0.87 &0.66 &0.85 &0.92\\
                  &($M=150$ GeV)&0.59 &0.69 &0.85 &0.79 &0.83 &0.90\\
\hline
\end{tabular}
\end{center}
\label{rmh120}
\end{table}

\section{Conclusion}
In LRTH model, the scalar $\hat{S}$ is a natural candidate for WIMP
dark matter, and the Higgs boson mass is typically in the range of
160 - 180 GeV. Since the invisible decay $h\to \hat{S}\hat{S}$ can
affect other decay branching ratios, and also has a strong
correlation with the scattering on nucleon, we in this work focused
on the low mass region of  $\hat{S}$ so that the decay $h\to
\hat{S}\hat{S}$ can be open. We obtained the following observations:
(i) The cross section of $\hat{S}$ scattering on nucleon can
naturally satisfy the CDMS II upper bound, and can be large enough
to be accessible at SuperCDMS; (ii) The Higgs boson can have a
sizable invisible decay $h\to \hat{S}\hat{S}$, whose branching ratio
can reach $32\%$ and has a strong correlation with the cross section
of $\hat{S}$ scattering on nucleon.  However, the branching ratios
of other new decay modes $h\to \hat{A}\hat{A}$ and $h\to
\hat{h}_1\hat{h}_1$ are small; (iii) The branching ratios of the
conventional decay modes of the Higgs boson, $h\to VV$ $(V=W,~Z)$
and $h\to b\bar{b}$, can be suppressed over $30\%$ and $47.5\%$,
respectively; (iV) The Higgs production cross sections times the
branching ratios of the conventional decays can be all sizably
suppressed. So, it is possible to probe the LRTH model via the Higgs
productions at the LHC.

\section*{Acknowledgment}
We thank Wenyu Wang for discussions and Shufang Su for providing
the Calchep Model Code. This work was supported in part by the
Foundation of Yantai University under Grant No.WL09B31, by the
National Natural Science Foundation of China (NNSFC) under grant
Nos. 10821504, 10725526 and 10635030, by the Project of Knowledge
Innovation Program (PKIP) of Chinese Academy of Sciences under
grant No. KJCX2.YW.W10 and by an invitation fellowship of
Tohoku University (Global COE of Japan).

\end{document}